\documentclass[conference]{IEEEtran}
\IEEEoverridecommandlockouts
\usepackage{cite}
\usepackage{amsmath,amssymb,amsfonts}
\usepackage[normalem]{ulem}
\usepackage{algorithm}
\usepackage{algpseudocode}
\usepackage{amsmath}
\usepackage{subfig}
\usepackage{xspace}
\usepackage{subfig}
\usepackage{graphicx}
\usepackage{multirow}
\usepackage{amsmath}
\usepackage{algorithm}
\usepackage{algpseudocode}
\DeclareMathOperator{\mlp}{MLP}

\newcommand{\modelname}{PTF-FedRec\xspace} 
\newcommand{\modelnamenospace}{PTF-FedRec}

\usepackage{graphicx}
\usepackage{textcomp}
\usepackage{xcolor}
\usepackage{multirow}
\def\BibTeX{{\rm B\kern-.05em{\sc i\kern-.025em b}\kern-.08em
    T\kern-.1667em\lower.7ex\hbox{E}\kern-.125emX}}
\begin{document}

\title{Hide Your Model: A Parameter Transmission-free Federated Recommender System
}

\author
{
	Wei Yuan{\small$^1\dag$}\hspace*{6pt}
    Chaoqun Yang{\small$^2\dag$}\hspace*{6pt}\thanks{$\dag$ Equal contribution.}
        Liang Qu{\small$^1$}\hspace*{6pt}
        Quoc Viet Hung Nguyen{\small$^2$}\hspace*{6pt}
        Jianxin Li{\small$^3$}\hspace*{6pt}
	Hongzhi Yin{\small$^{1*}$}\thanks{* Corresponding author.}
 % {$^*$}
 \hspace*{10pt}
  \\
	\fontsize{10}{10}\selectfont\itshape $~^1$The University of Queensland,
        \fontsize{10}{10}\selectfont\itshape$~^2$Griffith University,
        \fontsize{10}{10}\selectfont\itshape$~^3$Deakin University\\
	\fontsize{9}{9}\selectfont\ttfamily\upshape$~^1$\{w.yuan,l.qu1,h.yin1\}@uq.edu.au,\fontsize{9}{9}\selectfont\ttfamily\upshape$~^2$\{chaoqun.yang,henry.nguyen\}@griffith.edu.au\\\fontsize{9}{9}\selectfont\ttfamily\upshape$~^3$jianxin.li@deakin.edu.au
}

\maketitle

\begin{abstract}
  With the growing concerns regarding user data privacy, Federated Recommender System (FedRec) has garnered significant attention recently due to its privacy-preserving capabilities.
  Existing FedRecs generally adhere to a learning protocol in which a central server shares a global recommendation model with clients, and participants achieve collaborative learning by frequently communicating the model's public parameters.
  Nevertheless, this learning framework has two drawbacks that limit its practical usability:
  (1) It necessitates a global-sharing recommendation model; however, in real-world scenarios, information related to the recommendation model, including its algorithm and parameters, constitutes the platforms' intellectual property. 
  Hence, service providers are unlikely to release such information actively.
  (2) The communication costs of model parameter transmission are expensive since the model parameters are usually high-dimensional matrices. With the model size increasing, the communication burden will be the bottleneck for such traditional FedRecs.
  
Given the above limitations, this paper introduces a novel parameter transmission-free federated recommendation framework that balances the protection between users' data privacy and platforms' model privacy, namely \modelname.
  Unlike traditional FedRecs, participants in \modelname collaboratively exchange knowledge by sharing their predictions within a privacy-preserving mechanism. 
  Through this approach, the central server can learn a recommender model without disclosing its model parameters or accessing clients' raw data, preserving both the server's model privacy and users' data privacy.
  Besides, since clients and the central server only need to communicate prediction scores which are just a few real numbers, the communication overhead is significantly reduced compared to traditional FedRecs. 
  Extensive experiments conducted on three commonly used recommendation datasets with three recommendation models demonstrate the effectiveness, efficiency, and generalization of our proposed federated recommendation framework.
\end{abstract}


\section{Introduction}\label{sec:introduction}
As an effective solution to mitigate information overload by delivering personalized content to users from a large volume of data, recommender systems have been widely deployed in numerous web services, such as e-commerce~\cite{schafer2001commerce,zhang2021double}, social media~\cite{zhou2019online}, and online news~\cite{wu2020mind}.
Conventionally, a service provider designs a recommender and trains it on a central server using collected user raw data (e.g., user profiles and user-item interactions)~\cite{wang2016spore}.
However, this training approach poses significant risks of data leakage and privacy concerns~\cite{batmaz2019review}.
Given the recent release of privacy protection regulations in various countries and regions, like CCPA~\cite{harding2019understanding} in the USA, PIPL~\cite{calzada2022citizens} in China, and GDPR~\cite{voigt2017eu} in the EU, it has become increasingly challenging for online platforms to train a recommender using the traditional centralized training paradigm without violating these regulations.

Federated learning is a privacy-preserving learning scheme, in which clients can collaboratively learn a model without sharing their private data.
Therefore, recent studies attempt to utilize federated learning to train recommendation models, a.k.a, Federated Recommender Systems (FedRecs)~\cite{yang2020federated}.
~\cite{ammad2019federated} is the first FedRec framework that combines federated learning with a collaborative filtering model.
After that, many extended versions have been developed in a short time due to FedRec's privacy-preserving advantages~\cite{lin2020fedrec,chai2020secure,yi2021efficient,liu2022federated,guo2021prefer}.
% For example, Lin et al.~\cite{lin2020fedrec} investigated how to leverage explicit feedback in FedRecs.
% Chai et al.~\cite{chai2020secure} proposed a secure matrix factorization in federated scenarios.
% ~\cite{yi2021efficient,liu2022federated,guo2021prefer} further transplanted FedRecs to diverse scenarios such as news recommendation, social recommendation, and POI recommendation, respectively.

\begin{figure*}[!htbp]
  \centering
  \includegraphics[width=1.\textwidth]{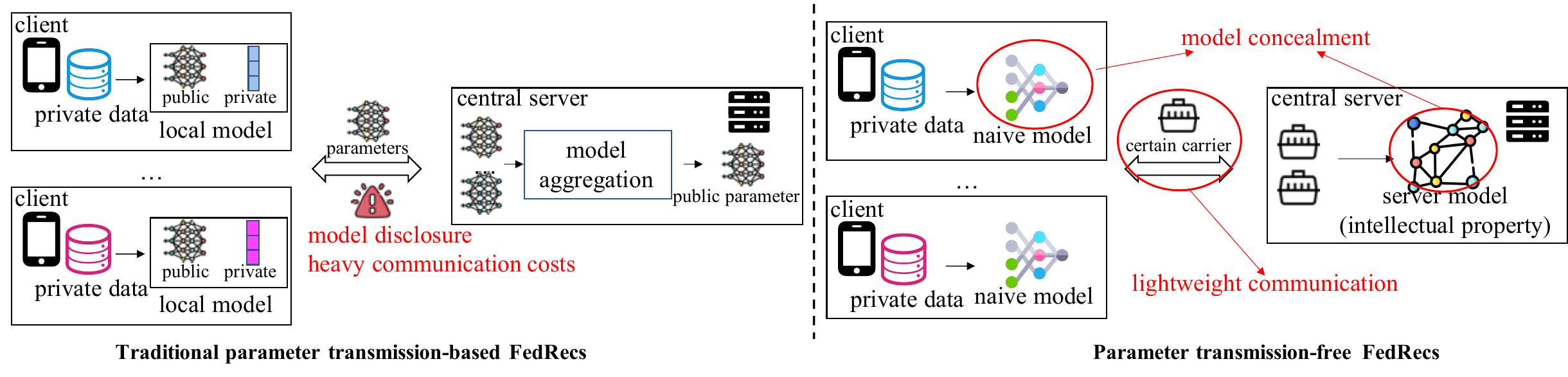}
  \caption{Traditional parameter transmission-based FedRec v.s. parameter transmission-free FedRec.}\label{fig_difference}
\end{figure*}

Despite the variety of proposed FedRecs, most of them follow a common learning protocol, where a central server coordinates clients to optimize a shared objective by transmitting and aggregating the parameters/gradients of a global recommender system~\cite{sun2022survey}, as shown in the left part of Fig.~\ref{fig_difference}.
To be convenient for presentation, we name these traditional FedRecs using the above learning protocol as parameter transmission-based FedRecs.
While these FedRecs offer a degree of protection for users' raw data, we contend that they are unsuitable for many real-world scenarios due to the following two main drawbacks.

The first limitation is that, these parameter transmission-based FedRecs require the central server to expose a recommendation model for knowledge-sharing purposes.
Unfortunately, in practical applications, especially within the commercial realm, the information pertaining to the recommendation models, including the design of model architecture and the value of model parameters, represents the core intellectual property of service providers.
Given the substantial expense involved in developing these models, few service providers are inclined to voluntarily disclose their models in the training process since competitors can easily plagiarize and re-distribute these valuable models by pretending to the normal users in federated recommender systems.
Unfortunately, all these parameter transmission-based FedRecs overlook the protection needs of the service providers' model privacy and even sacrifice the platform privacy to implement user privacy protection, dampening platforms' willingness to deploy these FedRecs.

Although some FedRec works~\cite{zhang2023comprehensive,chai2020secure} employ techniques like differential privacy to safeguard the recommendation model's public parameters, their original intentions are still for protecting user data from certain inference attacks, which cannot satisfy service providers' model privacy protection needs, as the model's architecture and optimization methods are still exposed to all participants.
In the realm of federated learning, some works explore to use the digital watermarking to protect intellectual property~\cite{tekgul2021waffle}.
Nevertheless, implementing digital watermarking in FedRecs presents significant challenges, primarily due to the considerably higher number of clients compared to traditional federated learning settings~\cite{yuan2023federated}.
Embedding such a large number of signatures can substantially impact model performance~\cite{yang2023federated}.
Furthermore, digital watermarking can only track model copying behavior but does not possess the capability to prevent it, so it is not a primary choice in real-world scenarios. 
As a result, ensuring model privacy in the context of parameter transmission-based FedRecs remains a formidable challenge, as their learning protocol inherently leaks model information.

Another shortcoming of current parameter transmission-based FedRecs is their huge communication expenses. 
Specifically, the public parameters of a recommendation model are frequently transmitted between clients and the central server to achieve collaborative learning.
These model parameters typically consist of high-dimensional matrices, leading to costly communication overhead. 
While some research efforts have put forth communication-efficient FedRecs~\cite{zhang2023lightfr}, their communication costs remain correlated with the size of the transmitted model.
With the increasing model size, the communication burden could potentially become a bottleneck for parameter transmission-based FedRecs in practical applications.

Generally, all the above-listed drawbacks of current FedRecs are due to using model parameters to transfer knowledge.
In light of this, a federated recommender system that does not need to disperse model parameters during collaborative learning, a.k.a., parameter transmission-free FedRec, is timely in demand.
As shown in the right part of Fig.~\ref{fig_difference}, in parameter transmission-free FedRecs, the central server's recommendation model is decoupled with clients' local models as they communicate via certain carriers unrelated to model parameters.
Therefore, the service provider can deploy an elaborately designed recommender model on the server side while assigning some straightforward and publicly available recommendation models on the client side, i.e., the model in the central server is hidden from clients since the server's and clients' models are heterogeneous.
Furthermore, if the carrier is more lightweight than the model parameters, the communication will be more efficient than traditional parameter transmission-based FedRecs.

It is worth noting that, although some federated learning studies~\cite{kulkarni2020survey} have delved into the investigation of model heterogeneity, they cannot be directly applied to build our parameter transmission-free FedRecs due to differing objectives.
Specifically, most of these studies primarily focus on achieving model diversity among clients to address resource imbalance issues.
These works either leverage a public proxy dataset to manage consensus~\cite{chang2019cronus,li2019fedmd,cho2022heterogeneous}, which, however, cannot be obtained in FedRecs as data samples in recommendation systems belong to specific users and sensitive. Or they still require transmitting model parameters between clients and the central server.
In contrast, the primary aim of our parameter transmission-free FedRec is to establish model heterogeneity between clients and the central server to protect the service provider's model intellectual property.
As a result, implementing a parameter transmission-free FedRec is non-trivial.

In this paper, we propose the first parameter transmission-free federated recommendation framework, named \modelname.
In \modelname, the central server and clients maintain distinct recommendation models.
As shown in knowledge distillation~\cite{hinton2015distilling}, the model's knowledge can be transferred via its prediction scores.
Therefore, in \modelname, the central server and clients communicate using their corresponding prediction scores.
More precisely, in each round, clients upload prediction scores for a subset of items.
To protect the client's private data, perturbations are introduced to clients' predictions.
The central server trains its model based on these uploaded predictions, as they collectively represent a form of collaborative information derived from different clients.
Subsequently, the central server provides broad collaborative information to clients by generating prediction scores for a set of high-confidence and hard negative items.
These steps are iteratively executed until model convergence is achieved.
To validate the effectiveness of \modelname, we conduct extensive experiments on three widely used recommendation datasets (MovieLens-100K~\cite{harper2015movielens}, Steam-200K~\cite{cheuque2019recommender}, and Gowalla~\cite{liang2016modeling}) using three recommendation models (NeuMF~\cite{he2017neural}, NGCF~\cite{wang2019neural}, and LightGCN~\cite{he2020lightgcn}).
The experimental results show that \modelname achieves better performance than parameter transmission-based FedRec baselines meanwhile obtains closer performance to the centralized training paradigm.
Further, the average experimental communication costs of \modelname for each client is $150$ to $2000$ times lower than commonly used FedRec baselines.

The main contributions of this paper are as follows:
\begin{itemize}
  \item To the best of our knowledge, we are the first to consider protecting the service provider's model privacy, i.e., the intellectual property of the model, in the context of federated recommender systems.
  \item We propose a parameter transmission-free federated recommendation framework, \modelname, which achieves federated collaborative learning via sharing prediction scores over of a subset of items. Compared to the current FedRec protocol, \modelname can balance the protection of both clients' data privacy and the service provider's model privacy, meanwhile, the communication expense of \modelname is also lightweight.
  \item Extensive experiments conducted on three public datasets with three recommendation models demonstrate the effectiveness, efficiency, and generalization of our methods. 
\end{itemize}

The remainder of this paper is organized as follows.
Section~\ref{sec_preliminary} provides the preliminaries related to our research, including the problem definition of federated recommender systems, the general learning protocol of current federated recommender systems, and the privacy protection demands in FedRecs.
Then, in Section~\ref{sec_methodology}, we present the technical details of our proposed federated recommendation framework.
The experimental results with comprehensive analysis are exhibited in Section~\ref{sec_experiments}, followed by the related works in Section~\ref{sec_related_work}.
Finally, Section~\ref{sec_conclusion} gives a brief conclusion of this paper.

\begin{table}[]
  \centering
  \caption{List of important notations.}\label{tb_notation}
  \begin{tabular}{l|l}
  \hline
   $\mathcal{D}_{i}$ & the local dataset for user $u_{i}$.  \\
   $\hat{\mathcal{D}}_{i}^{t}$ & the dataset created by user $u_{i}$'s local model in $t$ round. \\
   $\widetilde{\mathcal{D}}_{i}$ & the dataset created by server model for user $u_{i}$. \\
   \hline
   $\mathcal{U}$ & all users in the federated recommender system.  \\
   $\mathcal{U}^{t}$ & selected training users in $t$ round.  \\
   $\mathcal{V}$ & all items in the federated recommender system. \\
   $\mathcal{V}_{i}^{t}$ & trained items for user $u_{i}$ in $t$ round. \\
   $\hat{\mathcal{V}}_{i}^{t}$ & items selected to create dataset $\hat{\mathcal{D}}_{i}^{t}$. \\
   $\widetilde{\mathcal{V}}_{i}^{conf}$ & items selected based on confidence to create dataset $\widetilde{\mathcal{D}}_{i}^{t}$. \\
   $\widetilde{\mathcal{V}}_{i}^{hard}$ & hard negative items selected to create dataset $\widetilde{\mathcal{D}}_{i}^{t}$. \\
   \hline
   $r_{ij}$ & the preference score of user $u_{i}$ for item $v_{j}$.   \\ 
   $\hat{r}_{ij}$ & the predicted score for item $v_{j}$ by user $u_{i}$'s local model.   \\ 
   $\widetilde{r}_{ij}$ & the predicted score of $u_{i}$ for item $v_{j}$ by server model.   \\ 
   \hline
   $\mathbf{M}_{i}^{t}$ & user $u_{i}$'s model parameters in round $t$.   \\ 
   $\mathbf{M}_{s}^{t}$ & server model parameters in round $t$.   \\ 
   $\mathcal{F}_{c}$ & users' model algorithm.   \\ 
   $\mathcal{F}_{s}$ & server model algorithm.   \\ 
   \hline
   $\alpha$ & the size of server created dataset.   \\ 
   $\beta_{i}^{t}$ & the proportion of positive items selected to $\hat{\mathcal{D}}_{i}^{t}$.   \\
   $\gamma_{i}^{t}$ & the ratio of positive items and negative items in $\hat{\mathcal{D}}_{i}^{t}$.   \\ 
   $\lambda$ & the probability of swapping a positive item's scores.   \\ 
   \hline
  \end{tabular}
  \end{table}
\section{Preliminaries}\label{sec_preliminary}
% \textbf{Notations.} 
In this paper, bold lowercase (e.g., $\mathbf{a}$) represents vectors, bold uppercase (e.g., $\mathbf{A}$) indicates matrices, and the squiggle uppercase (e.g., $\mathcal{A}$) denotes sets or functions.
The important notations are listed in Table~\ref{tb_notation}.

\subsection{Problem Definition of Federated Recommender System}
% \textbf{General Federated Recommendation Settings.} 
Let $\mathcal{U}=\{u_{i}\}_{i=1}^{\left|\mathcal{U}\right|}$ and $\mathcal{V}=\{v_{j}\}_{j=1}^{\left|\mathcal{V}\right|}$ represent the sets of clients (users)\footnote{In this paper, the terms of ``client" and ``user'' are equivalent, since each client is solely responsible for one user.} and items, respectively.
$\left|\mathcal{U}\right|$ and $\left|\mathcal{V}\right|$ are numbers of clients and items.
In FedRec, each client $u_{i}$ manages its private dataset $\mathcal{D}_{i}$, which consists of the user's interaction records $(u_{i}, v_{j}, r_{ij})$.
$r_{ij}=1$ indicates that $u_{i}$ has interacted with item $v_{j}$, while $r_{ij}=0$ means $v_{j}$ is currently a negative item.
The goal of FedRec is to train a global recommender model that can predict each user's preference score for their non-interacted items, and then, select the top-K items with the highest prediction scores as recommendations. 

\subsection{Traditional Parameter Transmission-based Federated Recommender Systems}
Almost all existing FedRecs train a recommender model following the parameter transmission-based learning protocol~\cite{ammad2019federated,chai2020secure,lin2020meta,wu2022federated,lin2020fedrec,zhang2023comprehensive}.
In this protocol, a central server is required to open-source a recommendation model to all participants.
The open-source model is then divided into public parameters and private parameters.
The private parameters (typically user embeddings) are stored and maintained by corresponding clients, while the public parameters are transmitted between clients and the central server to collaboratively optimize the recommendation objectives with several global rounds.
Specifically, in round $t$, the central server first disperses public parameters to clients.
The clients combine received public parameters with their private parameters to form local recommender models. 
Subsequently, the clients train their local recommender model to optimize certain recommendation loss functions (e.g., BPRLoss~\cite{rendle2009bpr}) with a few local epochs.
After local training, the clients send the updated public parameters (or the gradients of the public parameters) back to the central server. 
Finally, the central server aggregates all received parameters using certain strategies (e.g., FedAvg~\cite{mcmahan2017communication}).
These above steps between the central server and clients are iteratively executed until model convergence.

\subsection{Privacy Protection in Federated Recommender Systems}\label{sec_privacy_fedrec}
In previous FedRec works~\cite{sun2022survey}, privacy protection mainly refers to protecting users' private data.
However, in practice, the recommendation models are the core intellectual entities and the service providers have involved multiple assets, including human expertise and computation resources, to develop these models.
Taking this into account, we argue that the privacy related to the central recommendation model is also critical.
Therefore, a privacy-preserving federated recommender system should satisfy both users' and service provider's privacy protection requirements.

Unfortunately, most existing parameter transmission-based FedRecs sacrifice the service providers' model privacy to protect users' data privacy, since their learning protocol requires the service provider to disclose its model architecture and optimization algorithm and transmit model parameters to all participants, impeding the practical usage.
Consequently, there exists a pressing need for a novel federated recommendation framework that can protect user privacy while simultaneously safeguarding the intellectual property (i.e., model privacy) of service providers.

\section{Methodology}\label{sec_methodology}
\begin{figure*}[!htbp]
  \centering
  \includegraphics[width=0.8\textwidth]{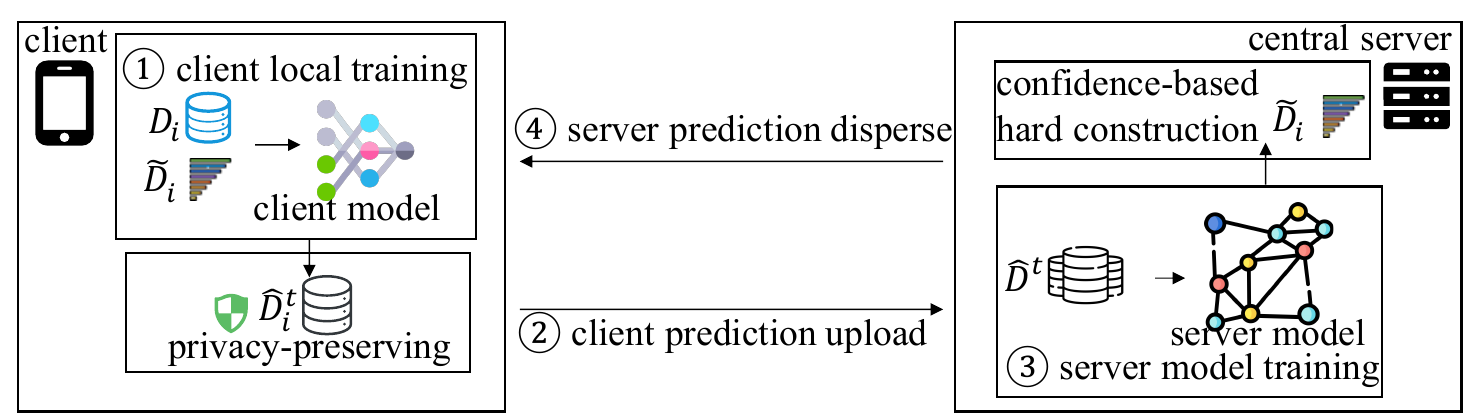}
  \caption{The details of \modelname.}\label{fig_ptf_fedrec}
\end{figure*}

The limitations of existing FedRecs discussed in Section~\ref{sec:introduction}, (1) overlook the model privacy protection and (2) have heavy communication costs, stemming from the utilization of model parameters for knowledge sharing.
Therefore, we propose a novel federated recommendation framework, \modelname, which does not rely on model parameters to achieve collaborative learning.
In this section, we first briefly introduce the basic recommendation models used in our framework, and then, we present the technical details of \modelname.
The overview of \modelname is illustrated in Fig.~\ref{fig_ptf_fedrec}.
Algorithm~\ref{alg_ours} presents \modelname's pseudo code. 

\subsection{Base Recommendation Models}\label{sec_basemodel}
Generally, a practical federated recommendation framework should be model-agnostic and compatible with most recommender systems. 
In this paper, to show the generalization of our proposed framework, we choose three popular recommendation models (NeuMF~\cite{he2017neural}, NGCF~\cite{wang2019neural}, and LightGCN~\cite{he2020lightgcn}) as our base models.
These three models comprehensively cover two main kinds of recommender systems: NeuMF for matrix factorization-based recommendation~\cite{mehta2017review}, while NGCF and LightGCN for graph-based recommendation~\cite{wu2022graph}.
The following is a brief introduction to these three models.

\textbf{Neural Matrix Factorization (NeuMF)}.
NeuMF~\cite{he2017neural} is a classic matrix factorization based recommender system. It leverages Multi-Layer Perceptron (MLP) and the concatenation of user and item feature vectors to predict the ratings:
\begin{equation}\label{eq_ncf}
  \hat{r}_{ij} = \sigma (\mathbf{h}^\top \mlp([\mathbf{u}_{i}, \mathbf{v}_{j}]))
\end{equation}
where $\mathbf{h}$ is trainable parameters and $[\cdot]$ is concatenation operation.

\textbf{Neural Graph Collaborative Filtering (NGCF) and LightGCN}.
NGCF~\cite{wang2019neural} and LightGCN~\cite{he2020lightgcn} are both graph-based recommender systems.
For them, users and items are treated as distinct nodes, and a bipartite graph is constructed according to the user-item interactions.
Then, user and item embeddings are computed by propagating their neighbor nodes' feature vectors, which can be generally formulated as follows:
\begin{equation}
  \begin{aligned}
    \mathbf{u}_{i}^{l} = \mathop{propagate}\nolimits^{l}(\mathbf{v}_{j}^{l-1};j\in\mathcal{N}_{u_{i}})\\
    \mathbf{v}_{j}^{l} = \mathop{propagate}\nolimits^{l}(\mathbf{u}_{i}^{l-1};j\in\mathcal{N}_{v_{j}})
    \end{aligned}
  \end{equation}
$\mathcal{N}_{u_{i}}$ and $\mathcal{N}_{v_{j}}$ are sets of neighbors for node $u_{i}$ and $v_{j}$, respectively. 
$propagate^{l}(\cdot)$ represents the propagation operation. $l$ represents the propagation layers. 
In NGCF, the propagation largely follows the standard GCN~\cite{kipf2016semi}, while LightGCN simplifies the NGCF's propagation by only keeping the neighborhood aggregation for training efficiency.
% \begin{scriptsize}
%   \begin{equation}
%     \begin{aligned}
%       \label{eq_ngcf}
%       \mathbf{u}_{i}^{l} = \sigma(\mathbf{W}_{1}\mathbf{u}_{i}^{l-1} + \sum\limits_{j\in \mathcal{N}_{u_{i}}}\frac{1}{\sqrt{\left| \mathcal{N}_{u_{i}} \right|} \sqrt{\left| \mathcal{N}_{v_{j}} \right|}} (\mathbf{W}_{1}\mathbf{v}_{j}^{l-1} + \mathbf{W}_{2}(\mathbf{v}_{j}^{l-1}\odot\mathbf{u}_{i}^{l-1}))) \\
%       \mathbf{v}_{j}^{l} = \sigma(\mathbf{W}_{1}\mathbf{v}_{j}^{l-1} + \sum\limits_{i\in \mathcal{N}_{v_{i}}}\frac{1}{\sqrt{\left| \mathcal{N}_{u_{i}} \right|} \sqrt{\left| \mathcal{N}_{v_{j}} \right|}} (\mathbf{W}_{1}\mathbf{u}_{i}^{l-1} + \mathbf{W}_{2}(\mathbf{u}_{i}^{l-1}\odot\mathbf{v}_{j}^{l-1})))
%       \end{aligned}
%     \end{equation}  
% \end{scriptsize}
% $\mathbf{W}_{1}$ and $\mathbf{W}_{2}$ are trainable parameters, $\odot$ is element-wise product.
% While for LightGCN, it simplifies E.q.~\ref{eq_ngcf} and only keeps the neighborhood aggregation for training efficiency and generation ability:
% \begin{equation}
%   \begin{aligned}
%     \label{eq_lightgcn}
%       \mathbf{u}_{i}^{l} = \sum\limits_{j\in \mathcal{N}_{u_{i}}}\frac{1}{\sqrt{\left| \mathcal{N}_{u_{i}} \right|} \sqrt{\left| \mathcal{N}_{v_{j}} \right|}}\mathbf{v}_{j}^{l-1}\\
%       \mathbf{v}_{j}^{l} = \sum\limits_{i\in \mathcal{N}_{v_{j}}}\frac{1}{\sqrt{\left| \mathcal{N}_{u_{i}} \right|} \sqrt{\left| \mathcal{N}_{v_{j}} \right|}}\mathbf{u}_{i}^{l-1}
%   \end{aligned}
%   \end{equation}
  After $L$ layers of propagation, the final user and item embeddings are used for predicting the ratings.
%   And the final user and item embeddings are used to predict $\hat{r}_{ij}$.

In \modelname, the clients and the central server possess different recommendation models. 
To simulate the setting that server model is elaborately designed while client models are simple and straightforward, we assume that the service provider assigns the simplest recommender model (i.e., NeuMF) to all clients.
And on the central server side, the model is not limited to the simple NeuMF, and it can employ more powerful recommendation models like NGCF and LightGCN. 
In Section~\ref{sec_explore_all_models}, we provide the performance results and analysis for all possible model combinations between client models and the server models.

\subsection{\modelname}
As mentioned in Section~\ref{sec_privacy_fedrec}, a privacy-preserving federated recommender system should safeguard both users' data privacy and service providers' model privacy.
Traditional FedRecs protect users' data privacy by hiding them on users' devices locally, however, they expose the service provider's recommendation model to clients.
As a result, these FedRecs essentially trade service provider's model privacy for users' data privacy, which discourages service providers from using them.
To address this challenge, we introduce a novel federated recommendation framework called \modelname. 
Unlike its predecessors, \modelname ensures the concealment of both the service provider's model and users' private data on their respective devices. 
Hence, the service provider's complex model is securely stored and maintained on a central server, while users' raw data remains on their individual devices. 
This approach guarantees the preservation of both model privacy and data privacy. 
To collaboratively optimize the model on the central server, \modelname employs an innovative learning protocol based on predictions.

\subsubsection{Prediction-based Learning Protocol}
Inspired by the knowledge transfer algorithms~\cite{bucilu2006model,hinton2015distilling}, in federated learning, some works~\cite{chang2019cronus,li2019fedmd,cho2022heterogeneous} have investigated training models by sharing predictions rather than transferring model weights.
Unfortunately, these works cannot be directly used in FedRecs to protect model privacy for the following reasons:
(1) Most of these methodologies rely on the creation of a public unlabeled dataset. In FedRecs, unlike the data samples $(x_{i},y_{i})$ widely used in the traditional federated learning tasks, the samples $(u_{i},v_{j},r_{ij})$ in the recommendation task are intricately linked to specific users and sensitive. Therefore, it is infeasible to construct a public shared recommendation sample set.
(2) The original aim of these methods is to foster model heterogeneity among clients rather than to protect the service provider's model privacy. In most of these studies, the central server is responsible solely for consolidating a consensus on the public dataset without hosting a model. 
Consequently, optimizing a model through prediction remains a non-trivial task in the context of FedRecs.
In this paper, we introduce a novel prediction-based learning protocol designed explicitly for FedRecs, catering to the privacy requirements of both clients and service providers.

The details of \modelname's learning protocol are as follows.

\textbf{Initial Stage.} Initially, the central server initializes an elaborately designed recommendation model (e.g., NGCF and LightGCN) on the server side, while clients initialize a simple recommendation model (e.g., NeuMF).
Subsequently, the central server and clients achieve collaborative learning by repeating the following four steps.

\textbf{Local Training on Client.}
In \modelname, at the global round $t$, a group of clients $\mathcal{U}^{t}$ are selected to participate in the training process. There are two datasets in client $u_{i}$'s local device: $\mathcal{D}_{i}$ and $\widetilde{\mathcal{D}}_{i}$.
The same as in conventional FedRecs, $\mathcal{D}_{i}$ is the client's private local dataset, which includes the user's positive and negative records $(u_{i},v_{j},r_{ij})$, $r_{ij}\in\{0,1\}$.
$\widetilde{\mathcal{D}}_{i}$ is the dataset obtained from the central server.
It contains triples with soft labels $(u_{i},v_{j},\widetilde{r}_{ij})$, $0\le\widetilde{r}_{ij}\le1$.
$\widetilde{r}_{ij}$ is the prediction score generated by the central server's model.
Note that if $u_{i}$ is the first time join training, $\widetilde{\mathcal{D}}_{i}=\emptyset$.

Let $\mathbf{M}_{i}^{t}$ denote the local recommendation model parameters for client $u_{i}$ in round $t$ and $\mathcal{F}_{c}$ is the recommendation algorithm. 
The client updates its local model on dataset $\mathcal{D}_{i}\cup\widetilde{\mathcal{D}}_{i}$ by optimizing the following loss function with several local epochs:
\begin{equation}
  \label{eq_client_objective}
  \begin{aligned}
    &\mathbf{M}_{i}^{t+1} = \mathop{argmin}\limits_{\mathbf{M}_{i}^{t}} \mathcal{L}^{c}(\mathcal{F}_{c}(\mathbf{M}_{i}^{t})|\mathcal{D}_{i}\cup\widetilde{\mathcal{D}}_{i})\\
    \mathcal{L}^{c} = -&\sum\nolimits_{(u_{i}, v_{j}, r_{ij})\in \mathcal{D}_{i}\cup\widetilde{\mathcal{D}}_{i}} r_{ij}\log \hat{r}_{ij} + (1-r_{ij})\log (1-\hat{r}_{ij})
  \end{aligned}
\end{equation}

\textbf{Client Prediction Upload.}
After local training, clients transfer their knowledge learned from dataset $\mathcal{D}_{i}\cup\widetilde{\mathcal{D}}_{i}$ back to the central server.
In \modelname, the knowledge is carried by the prediction results of the local model.
Specifically, the client $u_{i}$ first selects a group of items $\hat{\mathcal{V}}_{i}^{t}$, and then, it utilizes local model $\mathbf{M}_{i}^{t+1}$ to construct a prediction dataset $\hat{\mathcal{D}}_{i}^{t}$:
\begin{equation}
  \label{eq_simple_clientset}
  \begin{aligned}
    \hat{\mathcal{D}}_{i}^{t}&=\{(u_{i}, v_{j}, \hat{r}_{ij})\}_{v_{j}\in\hat{\mathcal{V}}_{i}^{t}}\\
    \hat{r}_{ij} &= \mathcal{F}_{c}(\mathbf{M}_{i}^{t+1}|(u_{i}, v_{j}))
  \end{aligned}
\end{equation}
Then, $\hat{\mathcal{D}}_{i}^{t}$ is uploaded to the central server to support the server's model training.
The construction of $\hat{\mathcal{D}}_{i}^{t}$ will directly influence the privacy protection of the client's data and the training performance of the central server model.
In Section~\ref{sec_client_data_upload}, we will present a privacy-preserving method for constructing $\hat{\mathcal{D}}_{i}^{t}$.

Algorithm~\ref{alg_ours} Lines 14-17 summarize the client's local training and client prediction uploading steps.

\textbf{Model Training on Server.}
The central server receives all prediction datasets $\{\hat{\mathcal{D}}_{i}^{t}\}_{u_{i}\in\mathcal{U}^{t}}$ and optimizes the following objective function:
\begin{equation}
  \label{eq_server_objective}
  \begin{aligned}
    &\mathbf{M}_{s}^{t+1} = \mathop{argmin}\limits_{\mathbf{M}_{s}^{t}}\sum\limits_{u_{i}\in \mathcal{U}^{t}} \mathcal{L}^{s}(\mathcal{F}_{s}(\mathbf{M}_{s}^{t})|\hat{\mathcal{D}}_{i}^{t})\\
    \mathcal{L}^{s} = -&\sum\nolimits_{(u_{i}, v_{j}, \hat{r}_{ij})\in \hat{\mathcal{D}}_{i}^{t}} \hat{r}_{ij}\log \widetilde{r}_{ij} + (1-\hat{r}_{ij})\log (1-\widetilde{r}_{ij})
  \end{aligned}
\end{equation}
where $\mathbf{M}_{s}^{t}$ is the central server's model parameters in round t, and $\mathcal{F}_{s}$ is the recommendation algorithm.

\textbf{Server Prediction Disperse.}
Since the server's model is trained on massive clients' uploaded predictions, it will achieve a more powerful recommender model.
Therefore, after the updating of the server model, the central server will disperse the learned knowledge back to clients to promote their local model training.
Specifically, for each client $u_{i}$, the central server selects a set of items $\widetilde{\mathcal{V}}_{i}$ and predicts the user's preference scores for these items using $\mathbf{M}_{s}^{t+1}$.
The predicted scores are transmitted to corresponding clients as dataset $\widetilde{\mathcal{D}}_{i}$:
\begin{equation}
  \label{eq_simple_serverset}
  \begin{aligned}
    \widetilde{\mathcal{D}}_{i}&=\{(u_{i}, v_{j}, \widetilde{r}_{ij})\}_{v_{j}\in\widetilde{\mathcal{V}}_{i}}\\
    \widetilde{r}_{ij} &= \mathcal{F}_{s}(\mathbf{M}_{s}^{t+1}|(u_{i}, v_{j}))
  \end{aligned}
\end{equation}
The effectiveness of server knowledge sharing depends on the selection of $\widetilde{\mathcal{V}}_{i}$.
In Section~\ref{sec_server_data_download}, we will further introduce a confidence-based hard knowledge dispersing method.
Algorithm~\ref{alg_ours} Lines 9-12 describe the server training and knowledge disperse steps.

The above is the basic learning protocol of \modelname.
The models on clients and the central server are collaboratively trained.
Specifically, the central server learns distributed knowledge via the prediction datasets $\hat{\mathcal{D}}_{i}^{t}$ uploaded by clients, meanwhile, clients augment local dataset based on $\widetilde{\mathcal{D}}_{i}$ generated by the central server.
In the following two subsections, we present the methods of constructing $\hat{\mathcal{D}}_{i}^{t}$ and $\widetilde{\mathcal{D}}_{i}$.

\subsubsection{Privacy-preserving $\hat{\mathcal{D}}_{i}^{t}$ Construction}\label{sec_client_data_upload}
Since the server's model is learned based on each client's uploaded dataset, the quality of $\hat{\mathcal{D}}_{i}^{t}$ will directly influence the performance of the server model.
To ensure the prediction quality, we restrict that the uploaded items in $\hat{\mathcal{D}}_{i}^{t}$ should be the trained items in client $u_{i}$, i.e., $\hat{\mathcal{V}}_{i}^{t}\subseteq \mathcal{V}_{i}^{t}$, since the prediction scores from non-trained items cannot provide any useful collaborative information.
Note that the trained item pool $\mathcal{V}_{i}^{t}$ consists of both positive and negative items, and the ratio of them is consistent with the predefined negative sampling ratio.

One naive way of developing $\hat{\mathcal{D}}_{i}^{t}$ is to upload predictions of the whole trained items, i.e., let $\hat{\mathcal{V}}_{i}^{t}=\mathcal{V}_{i}^{t}$. 
However, such a method will suffer privacy issues.
Specifically, assume the central server is curious but honest, that is, the central server is curious about clients' sensitive data (e.g., the positive items in our work) but it would not break the default learning protocol.
Then, if the client uploads the predictions of the whole trained items $\mathcal{V}_{i}^{t}$, the central server may be able to infer the client's interaction set by simply treating the items with top $\gamma\left|\mathcal{V}_{i}^{t}\right|$ prediction scores as the positive items.
$\gamma$ is the client's negative sampling ratio, which is often default set by the central server according to the best practice.
As the client's model parameters are optimized by E.q.~\ref{eq_client_objective}, the prediction scores of positive items have a large chance of being higher than negative items in the trained item set.
Therefore, the client's sensitive data will be leaked by such ``Top Guess Attack''. 

In traditional FedRecs, Local Differential Privacy (LDP)~\cite{park2022privacy} is widely used to protect user privacy.
However, LDP may be ineffective in this case as adding Laplace noise is still hard to change or conceal the order information of positive and negative items.
As a result, the central server still can infer users' interacted items via Top Guess Attack after applying LDP.

To safeguard users' data privacy, we design a privacy-preserving $\hat{\mathcal{D}}_{i}^{t}$ construction method, which includes two key steps: sampling and swapping.

\textbf{Sampling.} Inspired by the noise-free differential privacy~\cite{ijcai2021p216}, we only upload a subset of trained items. We protect each client's data privacy by concealing the positive and negative item ratio of the uploading dataset.  
Specifically, in round $t$, the client $u_{i}$ randomly initializes two values $\beta_{i}^{t}$ and $\gamma_{i}^{t}$.
$\beta_{i}^{t}$ is used to control the proportion of positive items that client $u_{i}$ will upload, while $\gamma_{i}^{t}$ controls the positive and negative item ratio for uploading.
For example, if $\beta_{i}^{t}$ is $0.1$ and $\gamma_{i}^{t}$ is $2$, the client $u_{i}$ will randomly select $10\%$ positive items and $2$ times size of negative items to form $\hat{\mathcal{V}}_{i}^{t}$.
Intuitively, since the ratio of positive items in the uploaded item sets is randomly changed, the central server cannot choose an appropriate $\gamma$ to execute the Top Guess Attack to obtain good inference results.
\begin{equation}
  \hat{\mathcal{V}}_{i}^{t} \leftarrow sample(\mathcal{V}_{i}^{t}|\beta_{i}^{t},\gamma_{i}^{t})
\end{equation}
After sample $\hat{\mathcal{V}}_{i}^{t}$, the client uses E.q.~\ref{eq_simple_clientset} to construct the prediction set $\hat{\mathcal{D}}_{i}^{t}$.

\textbf{Swapping.} Besides, to further protect user privacy, we propose a swap mechanism to perturb the client's uploaded predictions.
To be specific, the client randomly selects a proportion $\lambda$ of positive items with high prediction scores.
Subsequently, it exchanges these positive items' prediction scores with negative items.
\begin{equation}
  \hat{\mathcal{D}}_{i}^{t} \leftarrow swap(\hat{\mathcal{D}}_{i}^{t}|\lambda)
\end{equation}

\subsubsection{Confidence-based Hard $\widetilde{\mathcal{D}}_{i}$ Construction}\label{sec_server_data_download}
A client model with better performance can improve the server model's training, as the latter is trained using the prior one's predictions.
Therefore, in \modelname, at the end of each round, the central server will transfer the knowledge learned from the collective prediction data to each client via constructing and sharing the dataset $\widetilde{\mathcal{D}}_{i}$. 

Generally, a high-quality $\widetilde{\mathcal{D}}_{i}$ should have the following characteristics.
First, the knowledge conveyed via $\widetilde{\mathcal{D}}_{i}$ is ``reliable''.
Secondly, the transferred message is necessary for user $u_{i}$.
Based on these two requirements, we design a confidence-based hard sample construction method for \modelname.

\textbf{Confidence-based Selection.} 
To ensure the reliability of transferred knowledge, the central server sends items' predictions with high confidence to clients.
Intuitively, if an item embedding has been frequently updated, this item's embedding takes a large chance to be well-trained and the prediction calculated based on it will be closer to the truth.
Thus, in \modelname, we leverage the count of the server model's item embedding updates as the measure of the prediction's confidence to filter items.
Specifically, we first select items that have high update frequency and are not in client $u_{i}$'s uploaded dataset as the confidence-based selection item set $\widetilde{\mathcal{V}}_{i}^{conf}$.

\textbf{Hard Selection.} To ensure the necessity of transferred knowledge, the central server selects items with higher prediction scores for a client, as many works have demonstrated the positive impacts of hard negative samples~\cite{wu2021self,yuan2023federated}.

As a result, the central server selects items for a client $u_{i}$ formally as follows:
\begin{equation}\label{eq_widetilde_construct}
  \begin{aligned}
  &\widetilde{\mathcal{V}}_{i}^{conf} \leftarrow \mathop{argmax}\limits_{v_{i}\notin \hat{\mathcal{V}}_{i}^{t}\land \left|\widetilde{\mathcal{V}}_{i}^{conf}\right|=\mu*\alpha} frequency(\mathcal{V})\\
  &\widetilde{\mathcal{V}}_{i}^{hard} \leftarrow \mathop{argmax}\limits_{v_{j}\notin \hat{\mathcal{V}}_{i}^{t} \land \left|\widetilde{\mathcal{V}}_{i}^{hard}\right|=(1-\mu)\alpha} \mathcal{F}_{s}(\mathbf{M}_{s}^{t+1}|\mathcal{V})\\
  &\widetilde{\mathcal{V}}_{i} =  \widetilde{\mathcal{V}}_{i}^{conf} \cup \widetilde{\mathcal{V}}_{i}^{hard}
  \end{aligned}
\end{equation}
where $\alpha$ is the size of $\widetilde{\mathcal{D}}_{i}$ and $\mu$ controls the portion of confidence-based and hard selection.
These selected items $\widetilde{\mathcal{V}}_{i}$ are then used to construct $\widetilde{\mathcal{D}}_{i}$ according to E.q.~\ref{eq_simple_serverset}.

\begin{algorithm}[!ht]
  \renewcommand{\algorithmicrequire}{\textbf{Input:}}
  \renewcommand{\algorithmicensure}{\textbf{Output:}}
  \caption{\modelname} \label{alg_ours}
  \begin{algorithmic}[1]
    \Require global epoch $T$; local epoch $L$; learning rate $lr$, \dots
    \Ensure  server model $\mathbf{M}_{s}$
    \State server initializes model $\mathbf{M}_{s}^{0}$, clients initialize $\mathbf{M}_{i}^{0}$
    \State $\{\widetilde{\mathcal{D}}_{i}=\emptyset\}_{u_{i}\in\mathcal{U}}$
    \For {each round t =0, ..., $T-1$}
      % \State if $t$ is the epoch when the attack starts, $\mathcal{U} = \mathcal{U} \cup \widetilde{\mathcal{U}}$
      \State sample a fraction of clients $\mathcal{U}^{t}$ from $\mathcal{U}$
        \For{$u_{i}\in \mathcal{U}^{t}$ \textbf{in parallel}} 
        \State // execute on client sides
        \State $\hat{\mathcal{D}}_{i}^{t}\leftarrow$\Call{ClientTrain}{$u_{i}$, $\widetilde{\mathcal{D}}_{i}$}
        \EndFor
      % \State if $\widetilde{\mathcal{U}}_{t-1}$ is not empty, execute attack algorithm~\ref{alg_attack}
      \State // execute on central server
      \State receive client prediction datasets $\{\hat{\mathcal{D}}_{i}^{t}\}_{u_{i}\in\mathcal{U}^{t}}$
      \State $\mathbf{M}_{s}^{t+1}\leftarrow$ update server model using E.q.~\ref{eq_server_objective}
      \State update $\{\widetilde{\mathcal{D}}_{i}\}_{u_{i}\in\mathcal{U}^{t}}$ according to Section~\ref{sec_server_data_download}
    \EndFor
    \Function{ClientTrain} {$u_{i}$, $\widetilde{\mathcal{D}}_{i}$}
    \State $\mathbf{M}_{i}^{t+1}\leftarrow$ update local model using E.q.~\ref{eq_client_objective}
    \State construct $\hat{\mathcal{D}}_{i}^{t}$ according to Section~\ref{sec_client_data_upload}
    \State \Return $\hat{\mathcal{D}}_{i}^{t}$
  \EndFunction
    \end{algorithmic}
\end{algorithm}

\subsection{Discussion}
In this part, we discuss our proposed federated recommendation framework, \modelname, from two aspects: privacy-preserving and communication efficiency.

\subsubsection{Privacy Preserving Discussion}\label{sec_pp_analysis}
According to Section~\ref{sec_privacy_fedrec}, a privacy-preserving FedRec should provide both model privacy and user data privacy protection, therefore, we discuss these two types of privacy in \modelname here.

\textbf{Server Model Privacy Preserving}.
Unlike previous parameter transmission-based FedRecs that expose the model to all clients, the service provider's elaborate model in \modelname is hidden in the central server and cannot be accessed by any other participants.
In other words, the information on the server model, including model architectures, algorithms, and parameters, becomes black-box in \modelname.
Although some works explore copying a black-box recommendation model~\cite{fan2021attacking,chen2022knowledge,zhang2021reverse,zhang2023defense}, they are impractical to be applied in \modelname since they require accessing a large number of user data and user recommendation lists.
As a result, \modelname provides a safe environment for model privacy protection.

\textbf{User Data Privacy Preserving}.
In \modelname, following traditional FedRecs, clients' raw data are always stored in their local devices and cannot be accessed by other participants in the whole process, ensuring the security of users' original data.
However, similar to the traditional FedRecs the central server can infer the user's private data via uploaded public parameters~\cite{yuan2023interaction}, \modelname may leak the user's private information via the uploaded predictions.
To improve privacy, \modelname utilizes a noise-free differential privacy~\cite{ijcai2021p216} (i.e., sampling) with a swapping mechanism to protect the user's raw data.
According to~\cite{ijcai2021p216}, sampling method satisifies $(\epsilon, \delta)$-differential privacy.
Based on the post-processing property of differential privacy, applying swapping on the sampled data also satisfies $(\epsilon, \delta)$-differential privacy.
Therefore, \modelname can provide reliable protection for user data.

\subsubsection{Communication Efficiency Discussion}\label{sec_communication_analysis}
For traditional parameter transmission-based FedRecs, the communication costs for each client in every round exhibit a positive correlation with the size of the model's public parameters. 
These public parameters encompass item embeddings, $\mathbf{V}$, and other parameters denoted as $\mathbf{\Theta}$.
The communication cost of these conventional FedRecs can be represented as $\zeta \times size(\mathbf{V}+\mathbf{\Theta})$, with $\zeta$ symbolizing the efficiency factor.
As these public parameters $\mathbf{V}$ and $\mathbf{\Theta}$ generally constitute high-dimensional matrices, the communication costs for these FedRecs tend to be exorbitant.
While numerous communication-efficient FedRecs have been proposed~\cite{zhang2023lightfr}, their effects are only to curtail $\zeta$, thus their communication expenses remain contingent upon the magnitude of the model parameters.
As the model increases in complexity, these expenses ultimately become unmanageable. 
In contrast, for \modelname, the communication overhead for each client in every round can be characterized by $size(\hat{\mathcal{D}}_{i}^{t})$.
Considering the data sparsity inherent to each client and the fact that each data sample essentially comprises three real numbers $(u_{i}, v_{i}, r_{ij})$, the cost will be much lower than traditional FedRecs.

\section{Experiments}\label{sec_experiments}
In this section, we conduct experiments to answer the following research questions (RQs):
\begin{itemize}
  \item \textbf{RQ1}. How effective is our \modelname compared to centralized and conventional federated counterparts in recommendation performance?
  \item \textbf{RQ2}. How efficient is our \modelname compared to conventional federated counterparts in communication costs?
  \item \textbf{RQ3}. How effective is the privacy-preserving $\hat{\mathcal{D}}_{i}^{t}$ construction in \modelname?
  % \item \textbf{RQ4}. How is the influence of hyperparameters of privacy-preserving $\hat{\mathcal{D}}_{i}^{t}$ construction?
  \item \textbf{RQ4}. How effective is the confidence-based and hard sampling method for prediction dataset construction $D_i$ in \modelname?
  % \item \textbf{RQ6}. How does the value of the hyperparameter in the confidence-based hard server prediction construction affect \modelname's performance?
\end{itemize}

\subsection{Datasets}
We employ three real-world datasets (MovieLens-100K~\cite{harper2015movielens}, Steam-200K~\cite{cheuque2019recommender}, and Gowalla~\cite{liang2016modeling}) from various domains (movie recommendation, video game recommendation, and location recommendation) to evaluate the performance of \modelname.
The statistics of datasets are shown in Table~\ref{tb_statistics}.
MovieLens-100K includes $100,000$ records between $943$ users and $1,682$ movies. 
Steam-200K contains $3,753$ users and $5,134$ video games with $114,713$ interactions.
Gowalla is the check-in dataset obtained from Gowalla and we use a 20-core setting where $8,392$ users share $391,238$ check-in records on $10,068$ locations. 
Following previous works~\cite{ammad2019federated,he2017neural,yuan2023interaction}, we transform all positive ratings to $r_{ij}=1$, and negative items are sampled from non-interacted items with $1:4$ ratio during the training process.
All three datasets are randomly split into training and test sets with the ratio of $8:2$ and the validation data are randomly sampled from the client's local training set.
\begin{table}[!htbp]
  \centering
  \caption{Statistics of three datasets used in our experiments.}\label{tb_statistics}
  \begin{tabular}{l|ccc}
  \hline
  \textbf{Dataset}        & \textbf{MovieLens-100K} & \textbf{Steam-200K} & \textbf{Gowalla} \\ \hline
  \textbf{\#Users}        & 943                     & 3,753               & 8,392           \\
  \textbf{\#Items}        & 1,682                   & 5,134               & 10,086           \\
  \textbf{\#Interactions} & 100,000                 & 114,713             & 391,238        \\
  \textbf{Avgerage Lengths} & 106                 & 31             & 46        \\
  \textbf{Density} & 6.30\%                & 0.59\%             & 0.46\%        \\ \hline
  \end{tabular}
  \end{table}

\subsection{Evaluation Metrics}
We adopt two widely used evaluation metrics Recall at rank 20 (Recall@20) and Normalized Discounted Cumulative Gain at rank 20 (NDCG@20) to measure the recommendation performance.
We calculate the metrics scores for all items that have not interacted with users.
For the privacy-preserving evaluation, we use F1 scores to measure the inference performance of ``Top Guess Attack''.

\subsection{Baselines}
We compare \modelname with six baselines including both centralized and federated recommendation methods.

\noindent\textbf{Centralized Recommendation Baselines.}
We utilize NeuMF~\cite{he2017neural}, NGCF~\cite{wang2019neural}, and LightGCN~\cite{he2020lightgcn} as centralized recommendation baselines.
Note that we also use these models in our \modelname.
Thus this comparison will directly show the performance gap between centralized training and our federated training.
The introduction of these three baselines can be referred to Section~\ref{sec_basemodel}.

\noindent\textbf{Federated Recommendation Baselines.} 
We select three widely used federated recommendation frameworks as our baselines.
\begin{itemize}
  \item \textbf{FCF}~\cite{ammad2019federated}. It is the first work that extends the collaborative filtering model to federated learning. 
  \item \textbf{FedMF}~\cite{chai2020secure}. It is another privacy-preserving FedRec based on secure matrix factorization. Specifically, it utilizes homomorphic encryption techniques to protect user-level privacy on a distributed matrix factorization.
  \item \textbf{MetaMF}~\cite{lin2020meta}. It learns a meta-network on the central server and uses the meta-network to generate private personalized item embeddings for each user.
\end{itemize}

\subsection{Hyper-parameter Settings}\label{sec_hyper_default}
For all recommendation models, the dimensions of user and item embeddings are set to $32$. 
For NeuMF, three feedforward layers with dimensions $64$, $32$, and $16$ are used to process the concatenated user and item embeddings.
For both NGCF and LightGCN, the graph convolution weights' dimension is the same as the embeddings' size.
Besides, three GCN and LightGCN propagation layers are adopted in NGCF and LightGCN, respectively.
$\alpha$ is set to $30$.
For each client, $\beta_{i}^{t}$ is randomly sampled between $0.1$ to $1.0$ and $\gamma_{i}^{t}$ is randomly sampled from $1$ to $4$.
$\lambda$ is set to $0.1$ and $\mu$ is $0.5$.
We utilize Adam~\cite{kingma2014adam} with $0.001$ learning rate as the optimizer.
The maximum global rounds are $20$.
At each round, all clients participate in the training process.
The local training epochs for clients and the central server are $5$ and $2$, respectively.
For the server model, the training batch size is set to $1024$, while for the client model, the batch size is $64$.
The baselines of FedRecs are reproduced based on their papers.

\begin{table*}[!htbp]
  \centering
  \caption{The recommendation performance of \modelname and baselines on three datasets. \modelnamenospace(X) represents that the central server utilizes model ``X'', meanwhile the clients utilize NeuMF by default. The best performance of centralized recommendation is highlighted with underline, while the best performance of FedRecs is indicated by bold.}\label{tb_main}
  \begin{tabular}{ll|cc|cc|cc}
  \hline
  \multicolumn{2}{c|}{\multirow{2}{*}{\textbf{Methods}}}                               & \multicolumn{2}{c|}{\textbf{MovieLens-100K}} & \multicolumn{2}{c|}{\textbf{Steam-200K}} & \multicolumn{2}{c}{\textbf{Gowalla}}  \\
  \multicolumn{2}{c|}{}                                           & \textbf{Recall@20}     & \textbf{NDCG@20}    & \textbf{Recall@20}   & \textbf{NDCG@20}  & \textbf{Recall@20} & \textbf{NDCG@20} \\ \hline
  \multicolumn{1}{l|}{\multirow{3}{*}{\textbf{Centralized Recs}}} & \textbf{NeuMF}    & 0.1357              & 0.1544           & 0.3033            & 0.2074         & 0.0214          & 0.0177        \\
  \multicolumn{1}{l|}{}                                           & \textbf{NGCF}     & \underline{0.1883}              & \underline{0.2045}           & \underline{0.3777}            & \underline{0.2674}         & 0.0420          & \underline{0.0334}        \\
  \multicolumn{1}{l|}{}                                           & \textbf{LightGCN} & 0.1794              & 0.1954           & 0.3708            & 0.2640         & \underline{0.0442}          & 0.0333        \\ \hline
  \multicolumn{1}{l|}{}   & \textbf{FCF}      & 0.1108              & 0.1241           & 0.2341            & 0.1524         & 0.0150          & 0.0093        \\
  \multicolumn{1}{l|}{\multirow{3}{*}{\textbf{Federated Recs}}}                                           & \textbf{FedMF}    & 0.1192              & 0.1351           & 0.2444            & 0.1543         & 0.0160          & 0.0108        \\  
  \multicolumn{1}{l|}{}                                           & \textbf{MetaMF}    & 0.1138              & 0.1301           & 0.2356            & 0.1397         & 0.0153          & 0.0112        \\  
  \multicolumn{1}{l|}{}                                           & \textbf{\modelnamenospace(NeuMF)}    & 0.1319              & 0.1482           & 0.2554            & 0.1571         & 0.0202          & 0.0159        \\  
  \multicolumn{1}{l|}{}                                           & \textbf{\modelnamenospace(NGCF)}     & \textbf{0.1623}              & \textbf{0.1775}           & \textbf{0.3484}            & \textbf{0.2306}         & \textbf{0.0345}          & \textbf{0.0268}        \\ 
  \multicolumn{1}{l|}{}                                           & \textbf{\modelnamenospace(LightGCN)}     & 0.1606              & 0.1739           & 0.3246            & 0.2186         & 0.0330          & 0.0261        \\ \hline
  \end{tabular}
  \end{table*}

  \begin{table}[!htbp]
    \centering
    \caption{The comparison of average communication costs per client for one round. The costs for \modelnamenospace(NeuMF), \modelnamenospace(NGCF), and \modelnamenospace(LightGCN) are the same, thus we report them as \modelname to avoid repetition. The most efficient costs are indicated by bold.}\label{tb_communication}
    \begin{tabular}{lccc}
    \hline
    \textbf{Methods} & \textbf{MovieLens-100K} & \textbf{Steam-200K} & \textbf{Gowalla} \\ \hline
    \textbf{FCF}     & 0.46MB                  & 1.31MB              & 2.59MB           \\
    \textbf{FedMF}   & 7.32MB                  & 20.98MB             & 41.43MB          \\
    \textbf{MetaMF}   & 0.54MB                  & 1.63MB             & 3.22MB          \\ 
    \textbf{\modelname}    & \textbf{3.02}KB                  & \textbf{1.21}KB              & \textbf{1.59}KB           \\ \hline
    \end{tabular}
    \end{table}

\subsection{Effectiveness of \modelname (RQ1)}
We validate the effectiveness of our \modelname on three datasets with six baselines.
The experimental results are shown in Table~\ref{tb_main}.
``\modelnamenospace(X)'' indicates that the central server uses model ``X'' while the clients' models are always the naive NeuMF.
From the results, we have the following observations.

First of all, the centralized recommender systems achieve better performance than all federated recommendations. 
This may be because of two reasons: 
(1) Centralized training paradigm can directly access all data, however, FedRecs rely on certain knowledge carriers to achieve collaborative learning; 
(2) The privacy protection mechanism in FedRecs unavoidably introduces additional noises and consumes the recommendation performance.

Secondly, our \modelname consistently obtains better performance than FedRec baselines on all three datasets with different server models.
Specifically, when the central server's model becomes stronger, \modelname has better performance.
For example, when the central server's models are NGCF and LightGCN, i.e., \modelnamenospace(NGCF) and \modelnamenospace(LightGCN), our FedRecs even outperform some centralized recommender systems, e.g., centralized NeuMF.
Besides, according to Table~\ref{tb_main}, \modelnamenospace(NGCF) achieves the best performance among all FedRecs.

Thirdly, by comparing the performance across datasets, we can find that the sparsity of the dataset can significantly influence the performance gap between FedRecs and centralized recommender systems.
For example, on the denser dataset, such as MovieLens-100K, the performance of \modelnamenospace(NeuMF), \modelnamenospace(NGCF), and \modelnamenospace(LightGCN) have close performance to their corresponding centralized version respectively.
While on the sparser dataset, such as Gowalla and Steam-200K, the performance gap between centralized recommender systems and all FedRecs becomes larger.

\subsection{Communication Efficiency of \modelname (RQ2)}
Aside from its effective performance, the efficient communication of \modelname stands out as another advantage compared to traditional parameter transmission-based FedRecs.
In Section~\ref{sec_communication_analysis}, we generally analyze the difference in communication costs between \modelname and parameter transmission-based FedRecs.
The experimental results depicting average communication costs per client for \modelname and FedRec baselines are presented in Table~\ref{tb_communication}. 
Evidently, the communication costs for our \modelname are notably lower than all FedRecs baselines, as the communication costs for all FedRec baselines are at the level of megabytes, while the expense of \modelname is only at the kilobyte level.
Specifically, FedMF grapples with a heavy communication burden primarily due to its encryption process that expands the dimensions of item embeddings.
In contrast, \modelname incurs communication costs of about $3$KB for MovieLens-100K and under $1.6$KB for Steam-200K and Gowalla, which are at least $2000$ times lower than FedMF and $150$ times lower than FCF and MetaMF.
Moreover, across datasets, it is observable that the communication burden of traditional FedRecs is positively correlated with the number of items, as the item count directly impacts the size of item embeddings.
Consequently, the costs for all three baselines escalate from MovieLens-100K to Gowalla.
On the other hand, the costs for our \modelname are predominantly influenced by the average length of interactions for each client. Due to the sparsity of data in user-item interactions, \modelname consistently maintains lightweight communication costs across all three datasets.

\begin{table*}[!htbp]
  \centering
  \caption{The F1 scores of Top Guess Attack and NGCF@20 of \modelnamenospace(NGCF) with privacy-preserving $\hat{\mathcal{D}}_{i}^{t}$ construction. Lower F1 scores imply better privacy protection. ``$\downarrow$'' means the lower value is better, while ``$\uparrow$'' indicates higher scores are better. The best performance is shown by bold.}\label{tb_privacy}
  \begin{tabular}{l|cccccc}
    \hline
    \textbf{}                       & \multicolumn{2}{c}{\textbf{MovieLens-100K}}      & \multicolumn{2}{c}{\textbf{Steam-200K}}          & \multicolumn{2}{c}{\textbf{Gowalla}}             \\
    \textbf{Methods}                & \textbf{F1 Score}$\downarrow$ & \textbf{NDCG@20} $\uparrow$ &  \textbf{F1 Score}$\downarrow$ & \textbf{NDCG@20}$\uparrow$  & \textbf{F1 Score}$\downarrow$ & \textbf{NDCG@20}$\uparrow$  \\ \hline
    \textbf{No Defense}             & 0.9836            & \textbf{0.1909}                  & 0.9838            & \textbf{0.2494}                   & 0.9710            & \textbf{0.0281}                   \\
    \textbf{LDP}                    & 0.5873            & 0.1503                  & 0.8423            & 0.2176                   & 0.6782                 & 0.0251                 \\
    \textbf{Sampling}               & 0.5171            & 0.1834                  & 0.4706            & 0.2409                   & 0.4944            & 0.0274                    \\
    \textbf{Sampling + Swapping}    & \textbf{0.4539}            & 0.1775                  & \textbf{0.4016}            & 0.2306                   & \textbf{0.4236}            & 0.0268                \\ \hline
    \end{tabular}
  \end{table*}

  \begin{table}[!htbp]
    \caption{The $\frac{\Delta F1}{\Delta NDCG}$ scores for each privacy-preserving methods. Higher values imply the method consumes fewer model performance to protect user data privacy.}\label{tb_F1_NDCG_delta}
    \resizebox{0.48\textwidth}{!}{
    \begin{tabular}{lccc}
    \hline
    \textbf{Methods}           & \textbf{MovieLens-100K} & \textbf{Steam-200K} & \textbf{Gowalla} \\ \hline
    \textbf{LDP}               & 9.7                     & 4.45                & 97.6             \\
    \textbf{Sampling}          & \textbf{62.2}           & \textbf{60.3}       & \textbf{680.8}   \\
    \textbf{Sampling+Swapping} & 39.5                    & 30.9                & 421.1            \\ \hline
    \end{tabular}
    }
    \end{table}
    \begin{figure*}[!htbp]
      \centering
      \subfloat[MovieLens-100K.]{\includegraphics[width=0.85\textwidth]{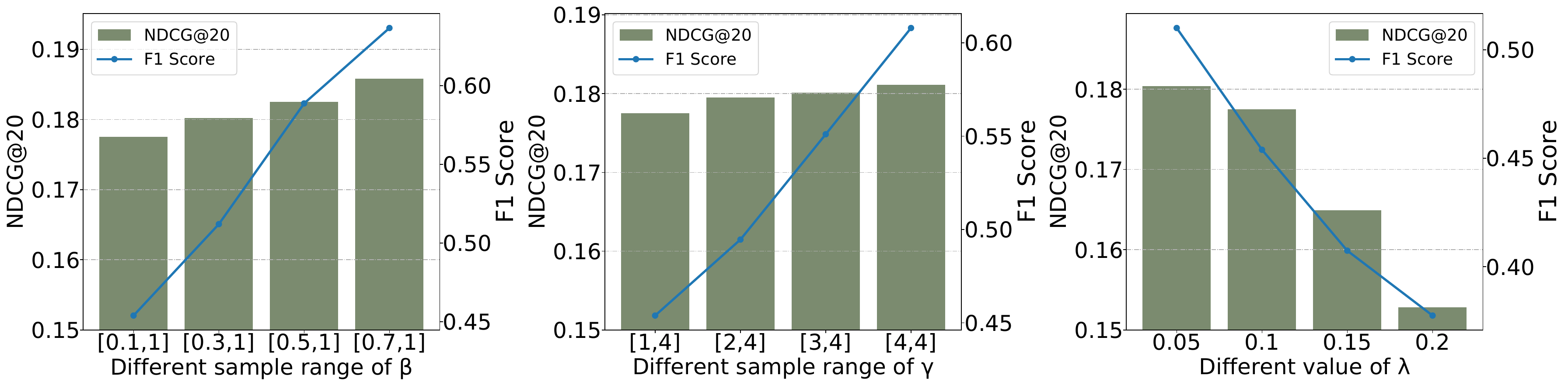}\label{fig_mlprivacy}}
      \hfil
      \subfloat[Steam-200K.]{\includegraphics[width=0.85\textwidth]{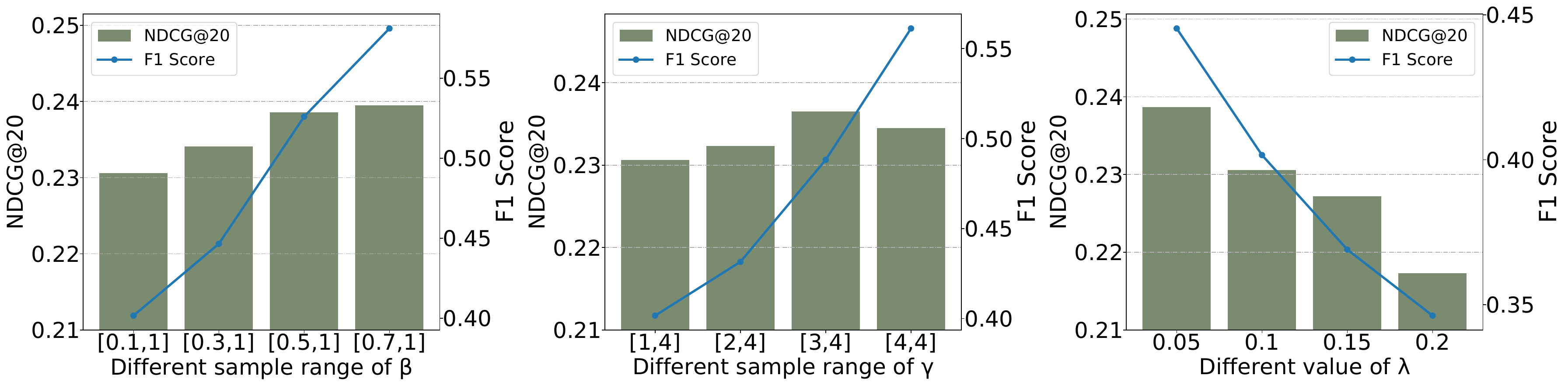}\label{fig_stprivacy}}
      \hfil
      \subfloat[Gowalla.]{\includegraphics[width=0.85\textwidth]{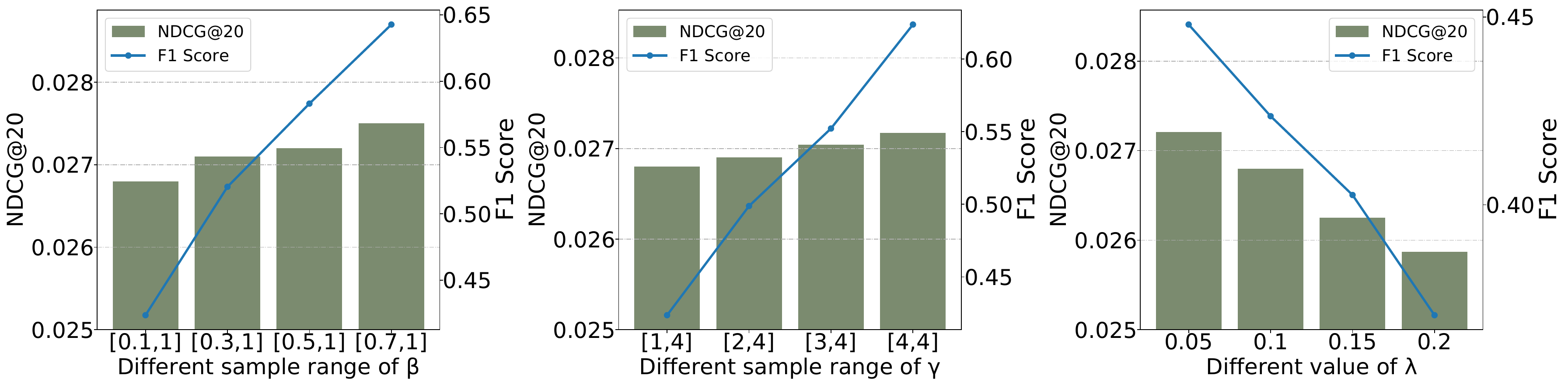}\label{fig_gwprivacy}}
      \caption{The impact of hyperparameter in privacy-preserving $\hat{\mathcal{D}}_{i}^{t}$ construction. $\beta_{i}^{t}$ controls the proportion of positive items that $u_{i}$ will upload, $\gamma_{i}^{t}$ regulates the ratio of negative items, $\lambda$ is the possibility of swapping a positive item's score.}\label{fig_privacy_hyperparameter}
    \end{figure*}

\subsection{Results of Privacy-preserving Mechanism (RQ3)}\label{sec_rq3}
In this section, we empirically showcase the effectiveness of privacy-preserving $\hat{\mathcal{D}}_{i}^{t}$ construction (Section~\ref{sec_privacy_effect_subsub}).
Then, we analyze the influence of hyperparameters in this privacy-preserving mechanism (Section~\ref{sec_hyp_privacy_subsub}).

To evaluate the privacy-preserving ability, the central server launches the ``Top Guess Attack'' mentioned in Section~\ref{sec_client_data_upload} for each client $u_{i}$'s uploaded predictions. 
That is, the central server guesses items with top $\gamma\left|\mathcal{V}_{i}^{t}\right|$ prediction scores as positive items.
In this paper, $\gamma$ is $0.2$ since the positive and negative item sampling ratio is $1:4$.
In Table~\ref{tb_privacy}, we present the attack's and recommender system's performance change after applying our privacy-preserving mechanism.
We compare our method with LDP, as LDP is the gold standard privacy protection method in traditional FedRecs.
Note that the privacy-preserving methods are unrelated to the server model type, therefore, we only show the results with \modelnamenospace(NGCF) by default, as it achieves the best model performance according to Table~\ref{tb_main}.

\begin{table*}[!htbp]
  \centering
  \caption{The impact of different item selection methods in $\widetilde{\mathcal{D}_{i}}$ construction for \modelname performance.}\label{tb_confidence_hard}
  \begin{tabular}{l|cccccc}
  \hline
                             & \multicolumn{2}{c}{\textbf{MovieLens-100K}} & \multicolumn{2}{c}{\textbf{Steam-200K}} & \multicolumn{2}{c}{\textbf{Gowalla}}  \\
  \textbf{Methods}           & \textbf{Recall@20}    & \textbf{NDCG@20}    & \textbf{Recall@20}  & \textbf{NDCG@20}  & \textbf{Recall@20} & \textbf{NDCG@20} \\ \hline
  \textbf{\modelname}              & \textbf{0.1623}                & \textbf{0.1775}              & \textbf{0.3484}              & \textbf{0.2306}            & \textbf{0.0345}             & \textbf{0.0268}           \\
  \textbf{-hard}                   & 0.1611                & 0.1724              & 0.3294              & 0.2126            & 0.0334             & 0.0262                \\
  \textbf{-confidence}             & 0.1602                & 0.1706              & 0.3256              & 0.2059            & 0.0323             & 0.0243                \\
  \textbf{-confidence -hard}       & 0.1566                & 0.1674              & 0.3107              & 0.1895            & 0.0316             & 0.0247                \\ \hline
  \end{tabular}
  \end{table*}
  \begin{figure*}[!ht]
    \centering
    \includegraphics[width=0.7\textwidth]{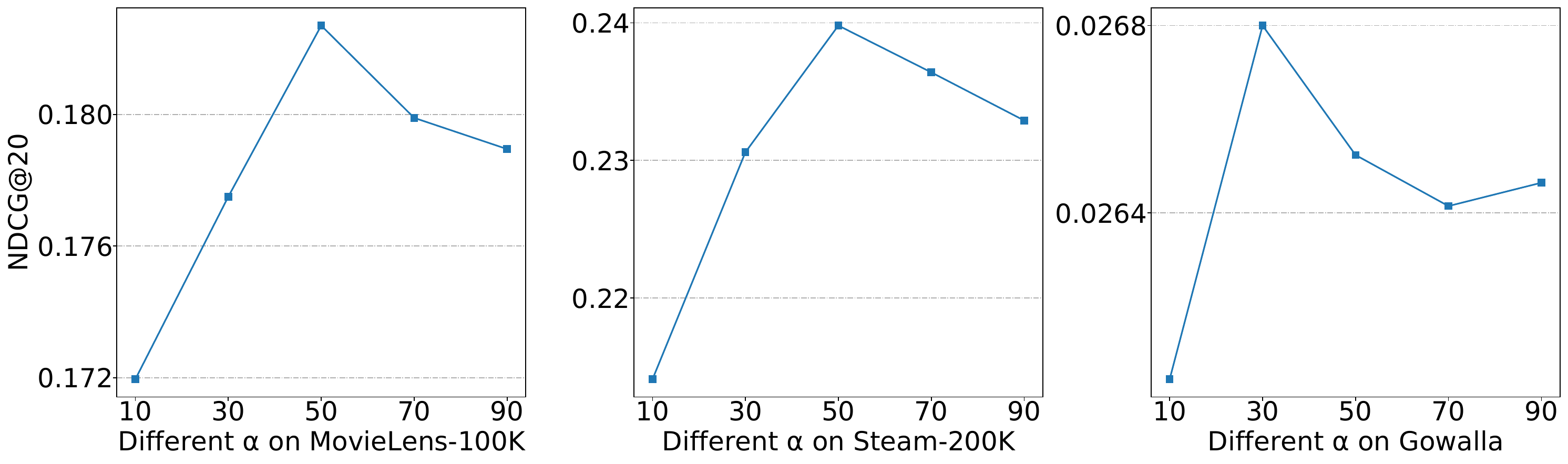}
    \caption{The impact of $\alpha$ (i.e., the size of server dispersed dataset $\widetilde{\mathcal{D}_{i}}$) on model performance.}\label{fig_serversize}
  \end{figure*}
\subsubsection{Effectiveness of Privacy-preserving $\hat{\mathcal{D}}_{i}^{t}$ Construction}\label{sec_privacy_effect_subsub}
According to the results in Table~\ref{tb_privacy}, when the client simply uploads all trained items' predictions to the central server, the curious server can obtain over $0.97$ F1 scores on all three datasets, which implies a severe data leakage of the positive items.
This is because the trained items' feature vectors are optimized by forcing positive items to have higher scores while negative items obtain lower scores, and the ratio of positive and negative items in the whole trained item set is assumed to be leaked to the central server.
To protect data privacy, LDP adds Laplace noise to the original prediction scores.
However, LDP may be ineffective in perturbing the order of prediction scores, and adding noise to all predictions will significantly reduce the utility of these prediction scores.
The results in Table~\ref{tb_privacy} also support this argument.
On MovieLens-100K, LDP reduces the attack's F1 scores from $0.98$ to $0.58$, but the recommender system's NDCG@20 scores are also decreased dramatically.
While for Steam-200K and Gowalla, the attack's performance still keeps around $0.8$ and $0.7$ F1 scores but the recommender system's performance is already compromised.

Unlike LDP, our \modelname protects the positive items by hiding the ratio of positive and negative items via sampling the uploaded dataset which will not sacrifice too much data utility.
Besides, to further protect the data privacy, \modelname adds ``noise'' to the uploaded prediction scores by swapping a small part of positive and negative items' scores, which can directly perturb the order information.
According to the results, when using sampling, the attack's F1 scores are reduced to around $0.5$ F1 scores on all three datasets.
When applying sampling and swapping defense methods, the attack's performance dramatically diminishes to about $0.4$ on all datasets.

Table~\ref{tb_F1_NDCG_delta} compares our defense methods with LDP by calculating the ratio of the attack's and the model's performance change ($\frac{\Delta F1}{\Delta NDCG}$). 
Higher scores indicate that the defense method safeguards data with less of a drop in model utility. 
According to the results, both Sampling and Sampling with Swapping are more cost-effective than LDP. 
It is noteworthy that although Sampling is more cost-effective than Sampling with Swapping, the latter can provide more powerful protection, as illustrated in Table~\ref{tb_privacy}. 
Therefore, the choice between using single Sampling or Sampling with Swapping depends on the privacy requirements of recommendation scenarios. 
If utility is prioritized, then only Sampling should be employed, whereas if privacy is more sensitive, Sampling with Swapping can be utilized.

% Our privacy protection methods are more cost-effective than LDP due to the multiple times higher $\frac{\Delta F1}{\Delta NDCG}$.

\subsubsection{Impact of Hyperparameters in Privacy-preserving Mechanism}\label{sec_hyp_privacy_subsub}
In \modelname's data protection method, there are three hyperparameters, $\beta_{i}^{t}$, $\gamma_{i}^{t}$, and $\lambda$.
Fig.~\ref{fig_privacy_hyperparameter} presents the result trends of these three hyperparameters with different settings.
Note that when we change one hyperparameter's value, the other two hyperparameters keep the default settings described in Section~\ref{sec_hyper_default}.

When we change the sampling range of $\beta_{i}^{t}$ from $[0.1, 1]$ to $[0.7, 1]$, the client is expected to select more positive items for the central server each round.
Therefore, both the model's performance and the attack's performance are increased.
For $\gamma_{i}^{t}$, when the sampling range changed from $[1, 4]$ to $[4, 4]$, the number of negative samples is expected to increase, meanwhile, the ratio of positive and negative items are becoming deterministic as the range shrunk.
Thus, the model performance is slightly improved while the attack's F1 scores are recovered dramatically.
Finally, we research the influence of $\lambda$ by changing its value from $0.05$ to $0.2$.
According to the right subfig of Fig.~\ref{fig_privacy_hyperparameter}, both attack and model performance are dropped with the growing of $\lambda$, since more proportion of positive items' prediction scores are swapped.

\subsection{Results of Confidence-based Hard $\widetilde{\mathcal{D}}_{i}$ Construction (RQ4)}\label{sec_rq4}
As the server model is trained on a lot of clients' uploaded predictions, it will capture broader collaborative information compared to clients' local models that are learned from clients' corresponding local data.
Enriching clients' knowledge with this more comprehensive collaborative information can indirectly improve the central server's model performance, as it is trained based on clients' uploaded predictions.
Therefore, in \modelname, at the end of each round, the central server constructs a dataset $\widetilde{\mathcal{D}}_{i}$ for each client $u_{i}$.
The items in $\widetilde{\mathcal{D}}_{i}$ are selected based on confidence and hardness strategies to ensure the reliability and necessity of shared information.
In this part, we first investigate the effectiveness of these item selection strategies (Section~\ref{sec_effect_conf_hard_subsub}). 
After that, we analyze dispersed dataset size's impact on model performance (Section~\ref{sec_data_size_subsub}).

\subsubsection{Effectiveness of Confidence-based Hard $\widetilde{\mathcal{D}}_{i}$ Construction}\label{sec_effect_conf_hard_subsub}
To validate the effectiveness of our confidence-based hard $\widetilde{\mathcal{D}}_{i}$ construction method, we gradually replace the confidence-based samples and hard samples with randomly selected items.
As shown in Table~\ref{tb_confidence_hard}, when we replace the hard samples (i.e., ``-hard'') or confidence-based samples (i.e., ``-confidence'') with random samples, the final model performance reduced from $0.1623$ to $0.1611$ and $0.1602$ Recall@20 scores respectively on MovieLens-100K.
Similar performance deterioration can also be found on Steam-200K and Gowalla datasets.
Furthermore, when we replace all the hard items and high confidence items with random samples (i.e., ``-confidence -hard''), the model performance further decreases to $0.1566$, $0.3107$, and $0.0316$ Recall@20 scores on three datasets respectively.
This phenomenon indicates both high-confidence items and hard items are more useful than randomly selecting a set of items' predictions for clients.

\subsubsection{Impact of $\widetilde{\mathcal{D}}_{i}$'s size}\label{sec_data_size_subsub}
We also explore the influence of different sizes of $\widetilde{\mathcal{D}}_{i}$ (i.e., the value of $\alpha$) for final model performance in Fig.~\ref{fig_serversize}.
Generally, when the value of $\alpha$ increases, the trend of performance of \modelname is at first increased to a peak point and then gradually decreased.
Specifically, on MovieLens-100K and Steam-200K, when $\alpha$ equals $50$, \modelname achieves the best performance, meanwhile, on Gowalla, the peak point is for $\alpha=30$.
This performance trend indicates that when the dispersed dataset is too small, the knowledge transferred from the server model to the client model is insufficient.
When the dispersed dataset is too large, the transferred knowledge may disturb client models learning from their own local datasets.

\subsection{Further Analysis}\label{sec_explore_all_models}
\begin{table}[!ht]
  \centering
  \caption{The performance (NDCG@20) of different model combinations for clients and the server on MovieLens-100K. Same observations can also be found on other two datasets.} \label{tb_all_combinations}
  \begin{tabular}{cl|ccc}
  \hline
                                                              &                   & \multicolumn{3}{c}{\textbf{Server Model}}          \\ \cline{3-5} 
                                                              &                   & \textbf{NeuMF} & \textbf{NGCF} & \textbf{LightGCN} \\ \hline
  \multicolumn{1}{c|}{\multirow{3}{*}{\textbf{Client Model}}} & \textbf{NeuMF}    & 0.1482         & \textbf{0.1775}        & 0.1739            \\
  \multicolumn{1}{c|}{}                                       & \textbf{NGCF}     & 0.1327         & 0.1711        & 0.1544            \\
  \multicolumn{1}{c|}{}                                       & \textbf{LightGCN} & 0.1386         & 0.1640        & 0.1549            \\ \hline
  \end{tabular}
  \end{table}

In the main experiments, we assume that clients utilize NeuMF and explore different models for the central server. 
In this section, we present the results of all model combinations for client and server models on MovieLens-100K in Table~\ref{tb_all_combinations}. 
Two interesting observations emerge from the results. Firstly, a more advanced server model yields better performance in horizontal comparison. 
Specifically, regardless of the client models used, the server model with NGCF exhibits the best performance, while the server model with NeuMF shows the worst performance. 
Secondly, a more complex client model leads to worse performance in vertical comparison; for instance, the client with NeuMF achieves the best performance regardless of the server model used. 
This outcome may be attributed to each client having limited data to support complex local model training due to data sparsity. 
Moreover, client local data can only construct a one-hop user-item graph. In contrast, graph-based recommender models such as NGCF and LightGCN are designed to capture high-order user-item relationships.

\section{Related Work}\label{sec_related_work}
\subsection{Federated Recommendation}
Federated recommender systems (FedRecs) have raised many researchers' interest recently due to their advantages of privacy protection~\cite{yin2024device}.
Ammand et al.~\cite{ammad2019federated} proposed the first federated recommendation framework with collaborative filtering models.
After that, many extended versions sprung up to improve the model performance~\cite{wu2022federated,lin2020fedrec,lin2020meta,chai2020secure,qu2023semi,zheng2023mmkgr,wang2021fast,nguyen2017argument,yuan2023hetefedrec} and transplanted FedRecs to various recommendation domains~\cite{yi2021efficient,liu2022federated,guo2021prefer,zheng2024decentralized,long2023model,ye2023heterogeneous}.
Besides, some works attempt to reduce the communication costs of FedRecs.
For example,~\cite{zhang2023lightfr} incorporated hash techniques to achieve lightweight communication, while~\cite{muhammad2020fedfast} proposed an active sampling method to accelerate the training process.
Given the achievements of FedRecs, the associated security concerns have been researched, such as the privacy issues~\cite{zhang2023comprehensive,yuan2023interaction,yuan2023federated,qu2024towards} and the robustness~\cite{zhang2022pipattack,yuan2023manipulating,yuan2023manipulating1}.

However, all these FedRecs are based on the parameter transmission-based learning protocol. 
As mentioned in Section~\ref{sec:introduction}, this learning protocol limits the practical usability of FedRecs as it overlooks the service providers' privacy needs and generates heavy communication costs.

\subsection{Model Heterogeneity in Federated Learning}
In federated learning, model heterogeneity has been introduced to alleviate resource imbalance problems, such as diverse data resource~\cite{ma2022state} and computation power disproportion~\cite{jiang2022fedmp,wang2022fedadmm}.
There are mainly two research lines to achieve model heterogeneity.
The first way is to design specific aggregation strategies based on target model architecture.
For instance,~\cite{diao2020heterofl,zhu2022resilient} proposed width-level strategies for different scales of CNN models' channel aggregation.
~\cite{wang2023flexifed,liu2022no} investigated layer-wise aggregation methods.
However, all these methods still rely on transmitting model parameters to fuse knowledge.

Another research line is to utilize predictions to transfer knowledge.
Specifically,~\cite{chang2019cronus,li2019fedmd,cho2022heterogeneous} proposed knowledge distillation-based federated learning framework.
In their works, a public reference dataset is built and clients transfer knowledge by making predictions on the public dataset.
The predictions are then aggregated on the central server to form ``consensus''.
Clients further update their local models based on the consensus.
These works are similar to our work that achieves collaborative learning based on model predictions, but there are still some differences: (1) their clients share a public dataset and upload predictions to achieve knowledge distillation, however, in \modelname, the prediction uploaded by clients are personalized and adaptive since public dataset is not available for FedRecs; (2) as their primary goal is to achieve client model heterogeneity, the central server in these works is mainly responsible for ``aggregate" client predictions, but the central server in \modelname aims to train its central server model to achieve model intellectual property protection. Other works, such as~\cite{he2020group,cho2022heterogeneous} not only use predictions but also clients' uploaded model parameters to achieve collaborative learning. As a result, the model heterogeneity methods in federated learning cannot be applied in federated recommender systems to protect service providers' model privacy.

\subsection{Model Privacy Protection in Federated Learning}
The model privacy includes two parts, model algorithm, and model parameters.
In federated learning, many works attempt to protect model parameters via differential privacy (DP) and encryption techniques~\cite{wei2020federated,park2022privacy,liu2022privacy}, but they overlook the leakage of model algorithms, such as model architectures.
Other works utilize watermarking techniques to protect the ownership of a model, however, these methods can only track the model copying behavior but cannot address the model leakage problem~\cite{tekgul2021waffle,lansari2023federated,yang2023federated}.
Therefore, the protection of the privacy of both model parameters and model architectures is still under-explored, especially in the context of federated recommender systems.

\section{Conclusion}\label{sec_conclusion}
In this paper, we propose a novel parameter transmission-free federated recommendation framework, \modelname, which achieves collaborative learning via transmitting predictions between clients and the central server.
In \modelname, the service provider does not need to expose its deliberate model, therefore, the model intellectual property has been protected.
Besides, since the dimension of predictions is much lower than recommendation model parameters, the communication costs of \modelname are much lighter than existing FedRecs.
To protect users' data privacy, \modelname incorporates a sampling and swapping mechanism for clients to share their local models' prediction scores.
A confidence-based hard sampling method is designed for the central server to disperse its learned collaborative knowledge.
Extensive experiments on three real-world recommendation datasets with three typical recommendation models demonstrate the effectiveness and efficiency of \modelname.

\section*{Acknowledgment}
This work is supported by the Australian Research Council under the streams of Future Fellowship (Grant No.FT210100624) and the Discovery Project (Grant No.DP240101108).

\bibliographystyle{IEEEtran}
\bibliography{IEEEabrv,IEEEexample}

\end{document}